----------------------------------------------------------------------------

\documentstyle [12pt,a4]{article}
\newcommand{\be}{\begin{equation}}
\newcommand{\ee}{\end{equation}}
\newcommand{\bea}{\begin{eqnarray}}
\newcommand{\ena}{\end{eqnarray}}
\newcommand{\vs}[1]{\rule[- #1 mm]{0mm}{#1 mm}}

\newcommand{\half}{\frac{1}{2}}

\newcommand{\slq}{U_q sl(2,{\bf C})}
\newcommand{\hsl}{U_q\widehat{sl}(2,{\bf C})}
\newcommand{\dof}{U_q\widehat{osp}(1,2)}
\newcommand{\dom}{U_qosp(1,2)}
\newcommand{\hr}{\hat{R}}
\newcommand{\vv}{v^{(1)}_0 \otimes v^{(2)}_0}
\newcommand{\vvv}{v^{(2)}_0 \otimes v^{(1)}_0}
\newcommand{\D}{\Delta}
\newcommand{\ot}{\otimes}
\newcommand{\wn}{w_0^n}
\newcommand{\wnm}{w_m^n}
\newcommand{\nn}{\nonumber}
\newcommand{\la}{\lambda}
\newcommand{\Ml}{M_\lambda}

\newcommand{\NP}[1]{Nucl.\ Phys.\ {\bf #1}}
\newcommand{\PL}[1]{Phys.\ Lett.\ {\bf #1}}
\newcommand{\CMP}[1]{Comm.\ Math.\ Phys.\ {\bf #1}}

\newcommand{\LMP}[1]{Lett.\ Math.\ Phys.\ {\bf #1}}

\begin{document}

\renewcommand{\thefootnote}{\fnsymbol{footnote}}
\newpage
\setcounter{page}{0}

\vs{15}

\begin{center}

{\large {\bf $R$-matrices for $\dof$ for highest weight  \\[.5cm]
representations of $\dom$ for general $q$ \\[.5cm]
and $q$ is an odd root of unity} \footnote{This work was supported
in part by the Bundesministerium fuer Forschung und Technologie}
}\\[1.5cm]

{\large T.S. Hakobyan \footnote{e-mail erphi@adonis.ias.msk.su}}\\
{\em Yerevan Physics Institute, Br.Alikhanian\\
st.2 , Yerevan 36, Armenia}\\[.5cm]

{\em and}\\[.5cm]

{\large A.G. Sedrakyan }\\
{\em Yerevan Physics Institute, Br. Alikhanian\\
st.2, Yerevan 36, Armenia}\\[.5cm]

\end{center}
\vs{10}

\centerline{\bf Abstract}

\indent
We obtain the formula for intertwining operator  ($R$-matrix) of
quantum universal enveloping superalgebra $\dof$ for $\dom$-Verma modules.
By its restriction we obtain the $R$-matrix for two semiperiodic (semicyclic),
two spin-$j$ and spin-$j$ and semiperiodic representations.

\vs{15}

\rightline{February 1993}

\vfill

\newpage
\renewcommand{\thefootnote}{\arabic{footnote}}
\setcounter{footnote}{0}
\newpage

\indent

 The intertwining operators of quantum universal enveloping algebras
\cite{Dr,J1,J2} lead to integrable 2D statistical systems. For $q$
(parameter of quantum group) isn't a root of unity they all are unified
in the Universal $R$-matrix for a given quantum group \cite{Dr,Rosso,T,T2}.
For $q$ is a root of unity its formal expression fails because of
singularities, arizing in this case. But intertwiners for cyclic
representations \cite{Date2,Arn1} exists in this case
also \cite{Bazh1,Bazh2,Date1,G1,G2,Iv} with a spectral parameter, lying
on some algebraic curve, and correspond to Chiral Potts model \cite{Bax,Y}.
In \cite{HS} using the method of \cite{G2} had constructed the general formula
for $R$-matrix of $\hsl$ for highest weight representations of $\slq$ both for
general $q$ and $q$ is a root of unity. \\
\indent
In this work we obtain its analogue for quantum superalgebra $\dof$
in a slightly different way. First we obtain the intertwiner for the tenzor
product of two Verma modules and then restrict it to other highest weight
representations of $\dom$. The graded Yang-Baxter equations are considered
also.

\indent

Quantum  superalgebra  $\dom$  is  generated   by   odd
elements $e, f$  and  even  element  $k$  with  (anti)  commutation
relations \cite{Saleur}:
\bea
kek^{-1} & =& q^\half e \qquad  k^{\pm1}k^{\mp1} = 1
\nonumber \\
kfk^{-1} & =& q^{-\half}f \label{eq:defrel} \\
{[}e,f{]}_+ &=&  \frac{k^2 -k^{-2}}{q-q^{-1}}:=\{k^2\}_q,
\nonumber
\ena
where $q$ is a deformation parameter.\\
\indent
In $\dom$ exists a Hoph algebra structure:
\bea
\D(e) & = & k \ot e+e\ot k^{-1} \qquad \D(k^{\pm1}) = k^{\pm1} \ot k^{\pm1}
\nonumber\\
\D(f) & = & k \ot f+f\ot k^{-1}
\label{eq:defcomul}
\ena
\indent
Recall that the multiplication in tensor product is defined by
the rule: $$(a\ot b)(c\ot d)=(-1)^{p(a)p(b)}ac\ot bd,$$  where $p(a)$
is  a parity of $a$. \\
\indent From (\ref{eq:defrel}) and (\ref{eq:defcomul}) it follows:
\be
\D (f)^n=\sum_{i=0}^n\left(
\begin{array}{c}
n \\
i
\end{array}
\right)_{-q}(k\ot f)^i(f\ot k^{-1})^{n-i},
\label{eq:Dpower}
\ee
where we use the notations
\bea
&& \left(\begin{array}{c} n \\m \end{array}\right)_x
:=\frac{(m)_x!}{(n)_x!(m-n)_x!}, \nn \\
&& (n)_x!:=(1)_x(2)_x\ldots (n)_x,
\qquad (n)_x:=\frac{x^n-1}{x-1} \nn
\ena
\indent
We  put  $k=q^h$, where $[h,e]=\half e$, $[h,f]=-\half f$.  From
(\ref{eq:defrel}) it follows (\cite{Saleur}):
\bea
 {[}e,f^n{]} = (-f)^{n-1}\frac{\left[ n\over 2\right]_+
     \left[2h-\frac{n-1}{2}\right]_+}{[1]_+}
\label{eq:power} \\
  {[}f,e^n{]} = (-e)^{n-1}\frac{\left[ n\over 2\right]_+
     \left[2h+\frac{n-1}{2}\right]_+}{[1]_+},
\nn
\ena
where we used the notation of \cite{Saleur}:
\[
    {[}x{]}_+ := \frac{q^x-(-1)^{2x}q^{-x}}{q^{1\over2}+q^{-{1\over2}}}
     \qquad (-1)^{2h}=1
\]
and  we  denote $[a,b]:=ab-(-1)^{p(a)p(b)}ba$ here and in the
following.\\
\indent
The center of $\dom$ is generated by $q$-deformed Casimir
element \cite{Saleur}
\be
   c=-(q^\half+q^{-\half})^2f^2e^2+(qk^2+q^{-1}k^{-2})fe+\{q^\half k^2\}_q^2
\label{eq:Casimir}
\ee
\indent
For $q$ not a root of unity the finite-dimentional  irreducible
representations $\pi_j$  are in one-to-one  correspondence  with  the
$\dom$ ones. They are characterized by nonnegative integer $j$ and
called spin-$\frac j2$ representations. The action of the $e,f,h$ in $\pi_j$
is given by \cite{Saleur}
\bea
&& \pi_j(h)v_m =q^{\half(j-m)}v_m \qquad
\pi_j(f)v_m =(-1)^mv_{m+1} \qquad \pi_j(f)v_{2j} =0 \nonumber \\
&& \pi_j(e)v_m =\frac{\left[m\over2\right]_+
                  \left[j+\frac{m-1}{2}\right]_+}{[1]_+}v_{m-1}
\label{eq:spinrep}
\ena
\indent
Consider now the case of $q$ is an odd root  of  unity.  Let  $N$  is  a
minimnal odd integer, satisfying the  condition  $q^N=1$.  It  follows
from (\ref{eq:defrel}) and (\ref{eq:power}) that then $e^{2N},f^{2N},k^{2N}$
belong to  the  center of $\dom$. In the irreducible representation
the values of these central elements  are  multiples  of  identity  operator.
Here we rewrite the action of $\dom$ on cyclic (periodic) representation,
which is characterized by 3 complex numbers $\la=q^{\half\mu},\alpha,\beta$
\cite{Sun}, in the following form
\bea
  \pi_{\la,\alpha,\beta}(f)v_m & = & v_{m+1} \qquad
       \pi_{\la,\alpha,\beta}(f)v_{2N-1} =\alpha v_0
 \nonumber \\
  \pi_{\la,\alpha,\beta}(e)v_m & = & \left[(-1)^{m-1}
      \frac{\left[m\over2\right]_+
       \left[\frac{2\mu-m+1}{2}\right]_+}{[1]_+}+(-1)^m
       \alpha\beta \right]v_{m-1} \label{eq:cyclrep} \\
  && (m=1\dots 2N-1), \nn \\
  \pi_{\la,\alpha,\beta}(e)v_0 & =&  \beta v_{2N-1} \qquad
     \pi_{\la,\alpha,\beta}(k)v_m =q^\frac{\mu-m}{2}v_m  \nn
\ena
The central elements $e^{2N}, f^{2N},k^{2N}$ take the values
\bea
 a:= \pi_{\la,\alpha,\beta}(e^{2N})=\beta\prod_{i=1}^{2N-1}
              (-1)^{i-1}\left[\frac{ \left[ \frac{i}{2} \right]_+
       \left[ \frac{2\mu-i+1}
       {2}\right]_+}
       {[1]_+}- \alpha \beta \right] \nn\\
 b:= \pi_{\la,\alpha,\beta}(f^{2N})=\alpha,
      \qquad \pi_{\la,\alpha,\beta}(k^{2N})=q^\mu
\ena
If $a$ or $b$ ($\alpha$ or $\beta$) vanishes  then the  cyclic
representation  converts into semiperiodic one, which is irreducible
for  generic  values of parameters also.\\
\indent
The   cyclic   and   semiperiodic   representations   have
no analoque for the classical $osp(1,2)$.

\indent

Now we will consider the affinization $\dof$ of $\dom$. It is generated by
elements $e_i,f_i,k_i$, $(i=0,1)$, which satisfy the following relations
\be
\begin{array}{ll}
  k_i^{\pm1}k_j^{\pm1} = k_j^{\pm1}k_i^{\pm1}, &
     k_i^{\pm1}k_j^{\mp1} = k_j^{\mp1}k_i^{\pm1} \ \ (i,j=0,1)  \\
  k_ie_jk_i^{-1} = q^\frac{a_{ij}}{2}e_j &
     k_if_jk_i^{-1} = q^{\frac{-a_{ij}}{2}}f_j \\
   {[}e_i,f_j{]}_+ = \delta_{ij} \{k_i^2\}_q, & \quad
\end{array}
\label{eq:defafrel1}
\ee
where $a_{ij} = \left( \begin{array}{rr}
1 & -1 \\
-1 & 1
\end{array}
\right)$ is Cartan matrix of $\dof$, and Serre relations. Here
$p(e_i)=p(f_i)=1, \quad p(k_i)=0$. \\
\indent
On $\dof$ there is a Hopf agebra structure:
\be
\begin{array}{ll}
\D(e_i)=e_i\ot k^{-1}_i+k_i\ot e_i &
\D(k_i^{\pm1})=k_i^{\pm1}\ot k_i^{\pm1} \\
\D(f_i)=f_i\ot k^{-1}_i+k_i\ot f_i &
\end{array}
\label{eq:defafcomul}
\ee
\indent
For any complex $x$ there is the homomorphism $\rho_x$:
$\dof\rightarrow\dom$:
\be
\begin{array}{lll}
\rho_x (e_0) = (-1)^pxf & \rho_x (f_0)=(-1)^px^{-1}e &
\rho_x (k^{\pm 1}_0) = k^{\mp 1} \\
\rho_x (e_1) = e & \rho_x (f_1) = f & \rho_x (k^{\pm 1}_1) = k^{\pm 1}
\end{array}
\label{eq:rho}
\ee
\indent
This homomorphism converts any representation of  $\dom$  to
parameterized representation  of  the  affine  quantum  algebra
$\dof$.  Using  $\rho_x$   we  can   construct   the   parametrised
representation  of $\dof$  from   the   representation   of
$\dom$.

\indent

The Hopf algebra structure allows  to  consider  the
action of $\dof$ on tensor products of representations.  Let
$\pi_1$  and $\pi_2$   are  representations  of $\dom$ on  $V_1$ and
$V_2$ respectively. Then, as it was mentioned above,
$\pi_i(x_i):=\pi_i\circ\rho_{x_i}$, $i=1,2$  are the representations
of $\dof$. Let we suppose  the equivalence  of $\pi_1(x_1)\ot\pi_2(x_2)$
and $\pi_2(x_2)\ot\pi_1(x_1)$ as an $\dof$-modules. This means the
existence of some intertwining operator $\hr_{{\pi_1}{\pi_2}}(x_1,x_2)$
from $V_1\ot V_2$  into  $V_2\ot V_1$ such that \cite{J1}
\be
 \hr_{\pi_1 \pi_2} (x_1,x_2) \pi_1 (x_1) \ot\pi_2 (x_2) (\D(g))
 = \pi_2 (x_2)\ot\pi_1 (x_1) (\D(g)) \hr_
{\pi_1 \pi_2} (x_1, x_2), \label{eq:inter}
\ee
where $g\in\dof$. We put $x_2=1, x_1=x$ because $\hr$ depends  on
$\frac{x_2}{x_1}$ only. Then the equations (\ref{eq:inter}) for $g=f_1,e_0$
can be represented in the following form:
\bea
\begin{array}{l}
\hr(x)(f\ot k^{-1}+k\ot f) = (f\ot k^{-1}+k\ot f)\hr(x)  \\
\hr(x)(x(-1)^pf\ot k+k^{-1}\ot (-1)^pf) =
      ((-1)^pf\ot k+xk^{-1}\ot (-1)^pf)\hr(x)
\end{array}
\label{eq:1efinter}
\ena
{}From (\ref{eq:1efinter}) we obtain:
\bea
\hr(x)(1\ot f) = [\D(f_1)\hr(x)x((-1)^{p+1}\ot k^2)-\D(e_0)\hr(x)]\nn \\
\qquad \times (x(-1)^{p+1}k\ot k^2-k^{-1}\ot (-1)^{p+1})^{p+1} \nn \\
\label{eq:2efinter} \\
\hr(x)(f\ot 1) = [\D (f_1)\hr (x)(k^{-2}\ot (-1)^{p-1})-\D (e_0)\hr(x) ]
\nn \\
\qquad \times (k^{-2}\ot (-1)^{p+1}k^{-1}-x(-1)^{p+1}\ot k))^{-1},\nn
\ena
where for simplicity we use the notations:
\[
\D (f_i):=\pi_2(1)\ot\pi_1(x)\left(\D (f_i)\right) \qquad
\D (e_i):= \pi_2(1)\ot \pi_1(x)\left(\D (e_i)\right).
\]
\indent
If $\pi_1$ and $\pi_2$ are the highest weight $\dom$-representations
with highest vectors $v_0^{(1)}$ and $v_0^{(2)}$ respectively and
$V_1\ot~V_2$ decomposes into direct sum of pairwise nonequivalent irreducible
representations, then we can normalize $\hr(x)$ such that
$\hr(x)(\vv)=\vvv$. In this case we can use (\ref{eq:2efinter}) to obtain
the recursive  formula  for  $\hr(x)(f^{r_1}v_0^{(1)}\ot{f^{r_2}}v_0^{(2)})$.
In fact, we obtain from (\ref{eq:2efinter}):
\bea
&&\hr(x) (f^{r_1}\ot f^{r_2})\vv= \nn\\
&& = (-1)^{r_1}\left[ x\lambda^2_{(2)}
(-1)^{p(v_0^{(1)})+r_1+1}q^{-r_2+1}\D(f_1)-\D(e_0)\right]\hr(x) (f^{r_1}\ot
f^{r_2-1})\times \nn \\
&& \times (q^{-r_2 -\frac{r_1}{2}+1}x(-1)^{p+1+r_1}
k\ot k^2-q^\frac{r_1}{2}(-1)^{r_2}k^{-1}\ot (-1)^p)^{-1} \vv \nn\\
\label{eq:3efinter} \\
&&\hr(x)  (f^{r_1}\ot f^{r_2})\vv=  \nn\\
&& =\left[\lambda^{-2}_{(1)}(-1)^
{p(v_0^{(2)})+r_2+1}q^{r_1-1}\D (f_1)-\D (e_0)\right]\hr(x)(f^{r_1-1}\ot
f^{r_2})\times \nn\\
&& \times ((-1)^{r_2}q^{r_1 +\frac{r_2}{2} -1}
k^{-2}\ot (-1)^{p+1}k^{-1} -q^{-\frac{r_2}{2}}x(-1)^p\ot k)^{-1} \vv \nn
\ena
Here $\lambda_{(i)} (i=1,2)$ are the values of $k$ on highest vectors
$v^{(i)}_0$.Using this we obtain by induction from (\ref{eq:3efinter}):
\bea
&&\hr(x) ( f^{r_1}\ot f^{r_2} ) \vv=  \nonumber \\
&&\qquad=\frac{(-1)^{r_1r_2}}
{\prod_{i=0}^{r_1 + r_2-1} \left( q^\frac{i}{2} (\lambda_{(1)}
\lambda_{(2)})^{-1}
(-1)^{p_1+p_2+i}-xq^{-\frac{i}{2}}\lambda_{(1)} \lambda_{(2)} \right)}\times
\nonumber \\
&&\qquad \times\prod^{r_2 -1}_{i_2 =0} \left[ x\lambda_{(2)}q^{-\frac{i_2}2}
(-1)^{p_1+i_1+1}\D(f_1) -\lambda^{-1}_{(2)}
q^{\frac{i_2}2} \D(e_0) \right]\times \label{eq:rmatrix} \\
&&\qquad \times\prod^{r_1 -1}_{i_1 =0} \left[ \lambda^{-1}_{(1)}
q^{\frac{i_1}2}
(-1)^{p_2} \D (f_1)-\lambda_{(1)}q^{-\frac{i_1}2} \D(e_0) \right]
 (\vvv) \nonumber
\ena

\indent

Let now $\pi_1$ and $\pi_2$ are Verma representations of $\dom$ on
$M_{\lambda_1}$ and $M_{\lambda_2}$ respectively. Then the elements
$\D(e_0)^n\D(f_1)^m\vv$ for all nonnegative integers $n$ and $m$ form the
basis of $M_{\la_1}\ot~M_{\la_2}$. From this and construction
procedure for $\hr(x)$ above follows the commutativity between
$\hr(x)$ and $\D(e_0),\ \D(f_1)$ on  $M_{\la_1}\ot M_{\la_2}$.
It's evident that $\hr(x)$ commutes with $\D(k_i)$, $(i=0,1)$ also.
We shall prove the commutativity between $\hr(x)$ and $\D(e_1),\ \D(f_0)$. \\
\indent
For general $q$ ($q$ isn't a root of unity) and $\lambda$
($\lambda\ne q^{\half j},\ j=0,1,\ldots)$ $M_\la$ is irreducibe.
For these values of $q$ and $\lambda_i$ the tenzor product of Verma modules
decomposes by the rule:
\be
M_{\lambda_1}\ot M_{\lambda_2}=\bigoplus_{n=0}^\infty M_{\lambda_1\lambda_2
q^{\half n}}
\ee
The highest vector $\wn$ of $M_{\lambda_1\lambda_2q^{\half n}}$ has the form:
\be
\wn=\sum_{j=0}^n c_j^i(\lambda_i,q) f^jv_0^{(1)}\ot f^{n-j}v_0^{(2)},
\label{eq:wn}
\ee
where the coefficients $c_j^n(\lambda_i,q)$ are some rational functions on
$\lambda_i$. (We can normalize $\wn$ in such a way that they would became
a polynomials on $\lambda_i$ and $\lambda_i^{-1}$). We assert that
\be
\D(e_1)\hr(x)\wn=0 \label{eq:1nulv}
\ee
\indent
Indeed, consider this equation for $\lambda_i=q^{\half l_i}$ for values of
$l_i$, which are large enough ($min(2l_1+1,2l_2+1)\geq n$).
For this values of $\lambda_i$ and general $q$
and also for spin-$\frac{l_i}2$ irreps (\ref{eq:spinrep}) $\hr(x)$ exists
and unique, as it had been shown in \cite{Saleur}. So $\hr(x)$ must map
the null vector of $V_{\lambda_1}\ot V_{\lambda_2}$ to the null vector of
$V_{\lambda_2}\ot V_{\lambda_1}$. Since $l_i$ are large enough,
the "border effects" of finite dimentionality of spin-$\frac{l_i}2$
irreps don't play the role in (\ref{eq:1nulv}). So, (\ref{eq:1nulv}) is
valid for infinite values of $\lambda_i$. Using expression (\ref{eq:rmatrix})
for $R$-matrix and (\ref{eq:wn}) we conclude that left hand side of
(\ref{eq:1nulv}) can be represented in the following form:
\be
\D(e_1)\hr(x)\wn=\sum_{j=0}^{n-1} b_j^i(\lambda_i,q) f^jv_0^{(1)}\ot
f^{n-j}v_0^{(2)},\label{eq:2nulv}
\ee
where the coefficients $b_j^n(\lambda_i,q)$ as $c_j^n(\lambda_i,q)$ above
are some rational functions on $\lambda_i$, that we can make
a polynomials on $\lambda_i$ and $\lambda_i^{-1}$ by choosing an appropriate
normalization of $\wn$. As they vanish for infinite many values of $\lambda_i$,
they vanish trivially. So, equation (\ref{eq:1nulv}) is valid. \\
\indent
Consider now the vectors $\wnm$, $\wnm=\D(f_1)^m\wn$ for any nonnegative
integers $n$ and $m$. They form a basis of $M_{\lambda_1}\ot M_{\lambda_2}$
 also. Note that $\hr(x)$ commutes with $\D(e_1)$. Indeed,
using (\ref{eq:1nulv}) and (\ref{eq:power}) and the commutativity between
$\hr(x)$ and $\D(f_1),\D(k_1)=\D(k)$ we obtain
\be
\begin{array}{l}
\D(e_1)\hr(x)\wnm=\D(e_1)\hr(x)\D(f_1)^m\wn=\D(e_1)\D(f_1)^m\hr(x)\wn=\\
\qquad =\D(f_1)^{m-1}\kappa(\D(k))\hr(x)\wn+
   (-1)^m\D(f_1)^m\D(e_1)\hr(x)\wn=\\
\qquad =\D(f_1)^{m-1}\hr(x)\kappa(\D(k))\wn=
   \hr(x)\D(f_1)^{m-1}\kappa(\D(k))\wn=\\
\qquad = \hr(x)\D(e_1)\D(f_1)^m\wn=\hr(x)\D(e_1)\wnm
\end{array}
\label{eq:prv}
\ee
where for simplisity we use the notation:
\[
\kappa_n(k)=\kappa_n(q^h):=
 (-1)^{n-1}\frac{\left[ n\over 2\right]_+\left[2h-\frac{n-1}{2}\right]_+}
 {[1]_+}
\]
So, ${[}D(e_1),\hr(x){]}=0$, because $\wnm$ form the basis of
$M_{\lambda_1}\ot M_{\lambda_2}$.\\
\indent
The same result may be obtained for $\D(f_0)$. To derive it we have to
consider the $\dom$-decomposition of $M_{\lambda_1}\ot M_{\lambda_2}$
with respect to $\dom$, generated by $\D(e_0),\D(f_0),\D(k_0)$. (Note
that in this case we deal with lowest weight Verma modules).
The basis vectors $\wnm$ must be replaced then by the basis vectors
$\tilde{\wnm}=\D(f_0)^m\tilde{\wn}$, where $\D(f_0)\tilde{\wn}=0$, and
\[
\D(k_0)\tilde{\wn}=(\lambda_1\lambda_2)^{-1}q^{\half n}\tilde{\wn}.
\]
The equation like (\ref{eq:prv}) may be written in this case also. \\
\indent
So, we proved that $\hr(x)$ is a $\dof$-intertwining operator:
\[
\hr(x):M_{\lambda_1}\ot M_{\lambda_2}\rightarrow{M_{\lambda_2}}
\ot{M_{\lambda_1}}.
\]
\indent
The graded Yang-Baxter equations on $M_{\lambda_1}\ot{M_{\lambda_2}}
\ot M_{\lambda_3}$
\be
(id\ot\hr(x))(\hr(xy)\ot id)(id\ot\hr(y)=
  (\hr(y)\ot id)(id\ot\hr(xy))(\hr(x)\ot id)
\label{eq:YB}
\ee
follows from the $\dof$-irreducibility of tenzor product, because both the
left and right sades of (\ref{eq:YB}) are $\dof$-intertwiners:
$M_{\lambda_1}\ot{M_{\lambda_2}}\ot{M_{\lambda_3}}\rightarrow{M_{\lambda_1}}
\ot M_{\lambda_2}\ot{M_{\lambda_3}}$ and both they map the
vector $v^{(1)}_0\ot v^{(2)}_0\ot v^{(3)}_0$ to the vector
$v^{(3)}_0\ot v^{(2)}_0\ot v^{(1)}_0$. \\
\indent
The above constucted $R$-matrix (\ref{eq:rmatrix}), intertwining Verma
modules, have the universality property in the following sence. Every highest
weight $\dom$-module $V_\la$ can be obtained by factorizing of $\Ml$ on some
submodule $I_\la$: $V_\la=\Ml / I_\la$. Consider the restriction of
equation (\ref{eq:inter}) on
\be
V_{\la_1}\ot V_{\la_2}=M_{\la_1} / I^1_{\la_1}\ot
M_{\la_2} / I^2_{\la_2}. \label{eq:fac}
\ee
If (\ref{eq:inter}) preserves the factorization, then $\hr(x)$ is well defined
on (\ref{eq:fac}). In a such way  $\hr_{V_1\ot V_2}(x)$ can be constructed
from $\hr_{M_{\la_1}\ot M_{\la_2}}$  by the restriction.

\indent

For instance, let us obtain an $R$-matrix for semiperiodic
representation $\pi_{\la,\alpha}:=\pi_{\la,\alpha,0}$ for $q^N=1$, where
$N$ is a minimal odd integer, satisfying this condition.
It characterized by two complex numbers $\la$ and $\alpha$
(See equations (\ref{eq:cyclrep}) for $\beta=0$). It is obtained from $\Ml$
by the factorization on the submodule, generated by the vector
$(f^{2N}-\alpha) v_0:  \quad V_{\la\alpha}=\Ml / I_\lambda^\alpha$. From
(\ref{eq:Dpower}) follows:
\bea
\D(f_1)^{2N}&=&k^{2N}\ot f^{2N}+f^{2N}\ot k^{-2N} \nn \\
\D(f_0)^{2N}&=&-k^{-2N}\ot e^{2N}-xe^{2N}\ot k^{2N}
\label{eq:2Npower}
\ena
\indent
The nesessary condition for constincensy of (\ref{eq:1efinter}) with
the factorization procedure is the coincidence of central elements
$\D(e_0)^{2N}$ and $\D(f_1)^{2N}$ in $V_{\la_1\alpha_1}\ot
V_{\la_2\alpha_2}$ and  $V_{\la_2\alpha_2}\ot V_{\la_1\alpha_1}$. Thus, from
(\ref{eq:2Npower}) follows
\be
\frac{\alpha_1}{\la_1^{2N}-\la_1^{-2N}}=
\frac{\alpha_2}{\la_2^{2N}-\la_2^{-2N}}, \qquad x^{2N}=1
\label{eq:cond}
\ee
But (\ref{eq:cond}) is a suffitient condition also. Indeed, from
(\ref{eq:2Npower}),(\ref{eq:1efinter}),(\ref{eq:cond}) follows
\be
\begin{array}{l}
\hr(x)\left[\la_1^{2N}(1\ot (f^{2N}-\alpha_2))+\la_2^{-2N}((f^{2N}-\alpha_1)
\ot 1\right]= \\
\qquad =\left[\la_2^{2N}(1\ot (f^{2N}-\alpha_1))+\la_1^{-2N}((f^{2N}-\alpha_2)
\ot 1\right]\hr(x) \\
\hr(x)\left[\la_1^{-2N}(1\ot (f^{2N}-\alpha_2))+x\la_2^{2N}((f^{2N}-\alpha_1)
\ot 1\right]\hr(x)= \\
\qquad =\left[x\la_2^{-2N}(1\ot (f^{2N}-\alpha_1))+\la_1^{2N}((f^{2N}-\alpha_2)
\ot 1\right]\hr(x)
\end{array}
\label{eq:ef2Ninter}
\ee
on  ${M_{\la_1}\ot M_{\la_2}}$. So, $\hr(x)(1\ot (f^{2N}-\alpha_2))$ or
$\hr(x)((f^{2N}-\alpha_1)\ot 1)$ can be represented as a linear combination
of $(1\ot (f^{2N}-\alpha_1))\hr(x)$ and $((f^{2N}-\alpha_2)\ot 1))\hr(x)$
As $f^{2N}$ is a central element of $\dof$ in $\Ml$, we finished the proof.\\
\indent
Note that (\ref{eq:cond}) for $x=1$ coincides with the like one for
$\slq$, derived in \cite{G1,G2}. \\
\indent
Obviously, the graded Yang-Baxter equations are satisfied on
$V_{\la_1,\alpha_1}\ot V_{\la_2,\alpha_2}\ot V_{\la_3,\alpha_3}$ for the
spectral parameter, lying on the algebraic curve
$$ \frac{\alpha_1}{\la_1^{2N}-\la_1^{-2N}}=
\frac{\alpha_2}{\la_2^{2N}-\la_2^{-2N}}=
\frac{\alpha_3}{\la_3^{2N}-\la_3^{-2N}},\qquad x^{2N}=1,\qquad y^{2N}=1 $$
\indent
The spectral parameter of $R$-matrix lies on the same algebraic curve
as in the case of $\slq$ \cite{G1,G2}.

\indent

In the case of $q^N=1$ we can consider also the $R$-matrix (\ref{eq:rmatrix})
for the mixed tensor products $V_{\la,\alpha}\ot V_j$ and
$V_j\ot V_{\la,\alpha}$ for $2j+1<2N$. Intertwiners of such type for $\slq$
and $U_qsl(n,{\bf C})$ had been considered in \cite{Arn2,Arn3,Arn4}. It can
be proven as above (see \cite{HS} for more detail for the case of $\slq$)
that the restriction of $\hr (x)_{\Ml\ot M_{q^{\half j}}}$
 ($\hr(x)_{M_{q^{\half j}}\ot \Ml}$) to $V_{\la,\alpha}\ot V_j$
($V_j\ot V_{\la,\alpha}$) is well defined for all values of parameters
$\la,\alpha,x$.\\
\indent
The graded Yang-Baxter equations are valid for mixed tenzor product of 3
representations because they are valid for corresponding Verma modules.
If there are 2 semiperiodic representations $V_{\la_1,\alpha_1}$ and
$V_{\la_2,\alpha_2}$, then their parameters must lie on the algebraic
curve (\ref{eq:cond})

\indent

Note that from any solution of graded Yang-Baxter equation the solution
of ordinary (nongraded) one can be obtained \cite{Ku}. The above considered
$R$-matrices of $\dom$ doesn't seem to give new solutions of Yang-Baxter
equations as for spin-$j$ irreps \cite{Saleur}. We suppose that they exhibit
hidden symmetries in Chiral Potts and mixed models.

\end{document}